\documentclass[12pt]{article}
\usepackage{amssymb}

\usepackage{amsmath}
\usepackage{epsfig}

\begin{document}

%--------+---------+---------+---------+---------+---------+---------
%Title page
\begin{titlepage}
\begin{center}
 \today \hfill UFR-HEP 00/23\\
\vskip 2cm
{\LARGE {\bf Solitons on Compact and Noncompact\\
~\\
   \bf Spaces in Large Noncommutativity}}
\vskip 2 cm
{\large El Mostapha Sahraoui  and El Hassan Saidi }
\vskip .8cm
   UFR-High Energy Physics, Department of Physics, Faculty of Sciences,\\
    Rabat University, Av. Ibn Battouta B.P. 1014, Morocco.
\vskip 0.5cm
{\tt H-saidi@fsr.ac.ma,  sahraoui@fsr.ac.ma} 

\vspace{2cm}
{\bf Abstract}
\end{center}

We study solutions at the minima of  scalar field potentials for
 Moyal spaces and  torii in the large non-commutativity and interprete
 these solitons in terms of non-BPS D-branes of string theory.
 We derive a mass spectrum formula linking different D-branes together  on  
quantum  torii and suggest that it  describes  general systems of D-brane bound
 states extending the D2-D0 one. Then
we propose a shape for the effective potential  approaching
 these quasi-stable bound states. We give the gauge symmetries of 
these systems of branes  and show that they depend  on the quantum
 torii representations.
\end{titlepage}

\newpage %%%%%%%

\section{Introduction}

It has been conjectured that at the stationary point of the tachyon
potential for non-BPS D-branes and for the D-brane-anti-D-brane pair of
string theories, the negative energy density cancels the brane tensions \cite{a,b,c}. 
This means that the minimum of the tachyon potential represents the
usual vacuum of the closed string theory without any D-branes. This
conjecture was studied in \cite{d} using a WZW-like open superstring field
theory free of contact term divergences where the condensation of the
tachyon field alone seems to give approximatively the vacuum energy. This
phenomenon was demonstrated directly in \cite{e} using Witten's string field
theory with cubic interaction \cite{f}.

Moreover tachyon condensation may also be studied using non-commutative
geometry (NC). This latter arises very naturally in string theory
when the antisymmetric background field is taken into account \cite{g,h,i,q}.
This important result has brought extra connections between the geometry of
D-branes and $K$-homologies on the $C^{\ast }$\ algebras and has opened new
issues in the analysis of non-commutative quantum field theory and string
field theory especially in the study of tachyon solitons. Indeed, it has
been shown that starting from D-branes of bosonic string theory and turning
on an antisymetric NS-NS $B$-field, we can get a condensation of tachyon
fields living on the world volume of the lower dimensional branes. The same
is true for non-BPS branes of type II superstrings and for the
D--brane-anti-D-brane systems. The main idea of this result is based on:
first, Sen's conjecture saying that it is not necessary to know exactly the
shape of the tachyon potential, what one really needs is its values at the
extrema. Second, the computation of the vaccum energy configurations where
the kinetic part of the effective action is neglected in front of the
potential term after non-commutative space coordinates rescaling.

On the other side, the scalar field is no more far to know the same
analysis. Indeed, it was shown \cite{j} that starting from the
non-commutative scalar field action, one cannot only proof the existence of
stable solitons but give them an approximate description at large
non-commutativity. This is not the end of the story since Harvey \textit{et
al} \cite{k,l} have given, based on the Gopakumar \textit{et al} (GMS) work 
\cite{j}, a more precise recipe for the solitonic solutions in terms of
lower dimentional D-branes whose dimensions depend intimately on the manner
of turning on the antisymetric NS-NS $B$-field.

All this material at hands, Gross and Nekrasov \cite{m,n,o} have developed
the notion of fluxon tubes in string theory by identifying gauge fields as
Higgs ones and minimising the energy of BPS D1-D3 system by solving the
Bogomoln'y and Nahm equations \cite{r,s,t}. Finally one ends with a magnetic
fluxon whose tension maches exacltly the fundamental string one.

In \cite{p}, Bars \textit{et al} have studied the tachyon condensation in
the non-commutative torus and predictd the existence of D$2$-D$0$ bound
states using Power-Rieffels projectors. We expect that this result is
generalised to higher dimensional compact quantum space involving a rich
spectrum of bound states. As a first step in this direction we will consider
higher dimentional torii.

The aim of this work is to study NC solitons in higher dimensional compact
and noncompact quantum spaces and explore, amongst others, the bound states
extending the Bars \textit{et al }solitons \cite{p}. Our analysis will be
made in two steps:

\begin{enumerate}
\item  Build NC solitons in $2l$ dimensional Moyal spaces, for $l\geq 1$.
This allows us to, first explore the features of NC solitons in higher
quantum spaces and the shape of the effective potential describing these
solutions. Second make an idea about the bound states one expects in $2l$
dimensional compact spaces. Finally, this part may be also extended to $%
\left( 2l+1\right) $\ quantum spaces where instead of D$0$-branes, one ends
up with electric fluxons; as such this part may be viewed also as an
extension of Harvey \textit{et al }analysis of D$25$-brane\cite{k}.

\item  Study of tachyon solitons for $2l$\ dimensional torii $\mathbb{T}%
_{\theta }^{2l}$\ and analyse the analogue of D$0$-D$2$ bound state
considered in \cite{p} for the quantum two torus. Since this later has
rational and irrational representations, we consider various realisations of 
$\mathbb{T}_{\theta }^{2l}$ and explore the corresponding bound states,
their mass spectrum and their underlying gauge symmetries.
\end{enumerate}

The paper is organised as follows: In section 2, we study non-commutative\
soliton solutions for scalar fields in Moyal plane $\mathbf{R}_{\theta }^{2}$
and generalise it to\ more wide non-commutative spaces. In section 3, we
analyse the non-commutative rational and irrational representations of torii
in order to give their respective projectors, the key for the tachyon
solutions. In section 4, we derive the exact formula of the mass spectrum of
the vaccum configurations and give an interpretation for its meaning. The
last section is devoted for discussions and conclude with the progress of
the present work.

\section{Non-commutative solitons on Moyal spaces}

In this section, we review the main lines of the NC soliton of a scalar
field theory on $\mathbf{R}_{\theta }^{2}\times \mathbf{R}$ and explore how
this kind of systems appears in low energy dynamics of string field theory.
Then we extend the corresponding results to non commutative scalar field
theories on $\mathbf{R}_{\mathbf{\theta }}^{2l}\times \mathbf{R}$ where now $%
\mathbf{R}_{\mathbf{\theta }}^{2l}$ is a $2l$ dimensional Moyal space whose
local coordinates $\left\{ x^{I}=\left( x^{2i-1},x^{2i}\right) ;\quad
i=1,2,\cdots ,l;\quad I=1,2,\cdots ,2l\right\} $ are taken such that 
\begin{eqnarray}
\left[ x^{2i-1},x^{2i}\right]  &=&\theta _{i},  \label{ncomm} \\
\left[ x^{2i\pm 1},x^{2j\pm 1}\right]  &=&0, \\
\left[ x^{2i},x^{2j}\right]  &=&0, \\
\left[ x^{2i\pm 1},x^{2j}\right]  &=&0,\quad i\neq j.
\end{eqnarray}
In eq (\ref{ncomm}) the $\theta _{i}$'s are $l$ real numbers which we choose
to be positive definite; otherwise one has just to rename the coordinate
variables and turning back to the first case. For example taking $\theta _{1}
$ negative and all the other $\theta _{i}$'s positive, one has just to set $%
x^{1}=x^{\prime 2}$ and $x^{2}=x^{\prime 1}$ and come back to the initial
case where all $\theta _{i}$'s are positive. As a matter of convention
notations, we shall use in what follows the convenient normalisation $\left(
y^{2i-1}=x^{2i-1}/\sqrt{\theta _{i}},\quad y^{2i}=x^{2i}/\sqrt{\theta _{i}}%
\right) $; but for commodity we shall continue to denote the $y^{2i-1}$\
and\ $y^{2i}$ variables as $x^{2i-1}$ and $x^{2i}$ respectively. Observe in
passing that for non zero $\theta _{i}$'s, $\mathbf{R}_{\mathbf{\theta }%
}^{2l}$ is a quantum space viewed as an algebra $\mathcal{A}_{\mathbf{\theta 
}}$ of endomorphisms of the hilbert space $\mathcal{H}$ of harmonic
oscillators; i.e\ $\mathcal{A}_{\theta }=End(\mathcal{H})$. To tie up this
discussion on the $\theta _{i}$'s, note also that they measure the non
commutativity of the space variables and have various interpretations.\ In
string theory, the $\theta _{i}\ $parameters are roughly speaking linked to
the inverse of the NS-NS $B$-field as 
\begin{eqnarray}
\theta ^{IJ} &=&-\left( 2\pi \alpha ^{\prime }\right) ^{2}\left( \frac{1}{%
g+2\pi \alpha ^{\prime }B}B\frac{1}{g-2\pi \alpha ^{\prime }B}\right) ^{IJ},
\\
G_{IJ} &=&g_{ij}-\left( 2\pi \alpha ^{\prime }\right) ^{2}\left(
Bg^{-1}B\right) _{IJ},
\end{eqnarray}
where $G_{IJ},$ and $g_{IJ}$, denote the effective open string and the
closed string metrics respectively. In the large noncommutativety these
equations reduce to 
\begin{eqnarray}
\theta ^{IJ} &=&\left\{ 
\begin{array}{ll}
\left( \frac{1}{B}\right) ^{IJ} & \qquad \quad i,j=1,\cdots ,2l \\ 
0 & \qquad \quad \text{otherwise}
\end{array}
\right.  \\
G_{IJ} &=&\left\{ 
\begin{array}{ll}
-\left( 2\pi \alpha ^{\prime }\right) ^{2}\left( Bg^{-1}B\right) _{IJ} & 
i,j=1,\cdots ,2l \\ 
g_{IJ} & \text{otherwise}
\end{array}
\right. .
\end{eqnarray}
In our present study, the $B$-field is taken as $B_{IJ}=B_{\left[ I/2\right]
}\Omega _{IJ}$ and similarly $\theta _{IJ}=\theta _{\left[ I/2\right]
}\Omega _{IJ},$ where $\Omega _{IJ}$ is the antisymmetric $2l\times 2l$
matrix of the symplectic form.

\subsection{Non-commutative soliton on \textbf{Moyal plane}}

We start by recalling that the field action $\mathcal{S}=\mathcal{S}(\phi )$
of a scalar field theory on the noncommutative $\mathbf{R}_{\theta
}^{2}\times \mathbf{R}$ space-time in the convention notation we are using
is: 
\begin{equation}
\mathcal{S=}\int_{\mathbf{R}_{\theta }^{2}\times \mathbf{R}}\text{d}^{3}%
\text{x}\left( \eta ^{\mu \nu }\partial _{\mu }\phi \partial _{\nu }\phi
+\theta V\left( \ast \phi \right) \right) .  \label{saction}
\end{equation}
Here $V\left( \ast \phi \right) $ is the potential of the noncommutative
field operator belonging to the noncommutative algebra $\mathcal{A}_{\theta }
$ introduced above. The $\ast $ product is the usual Moyal product
normalised to 
\begin{equation}
f(x)\ast g(x)=\text{exp}\left( \frac{i}{2}\epsilon _{IJ}\partial /\partial
x^{I}\partial /\partial y^{J}\right) \left[ f(x)g(y)\right] \mid _{x=y},
\label{star}
\end{equation}
where $\epsilon _{IJ}$ is the $2\times 2$ antisymmetric matrix. In eq (\ref
{star}), $f(x)$ and $g(x)$ are functions on the Moyal plane which by using
the Weyl correspondence may be interpreted as matrix operators $F$ and $G$
of the algebra\ $\mathcal{A=}End\mathcal{(H)}$ acting on the Hilbert space $%
\mathcal{H}$ of the harmonic oscillator. In this correspondence, $f\ast g$
is replaced by the usual matrix product $F.G$ and the integration with
respect to the non-commutative $x^{I}$'s is translated into the trace in $%
\mathcal{A}$. The scalar field operator $\phi (x)$ we consider here is
interpreted in string theory as the tachyon field $T(x)$ of the open string
ending on $D2$ brane. This may be immediatly seen in large non-commutativity
limit by comparing eq (\ref{saction}) to the following the tachyon effective
action $\mathcal{S}=\mathcal{S}(T(x))$ obtained from string field theory by
keeping $T(x)$ and integrating out all other fields 
\begin{equation}
\mathcal{S=}\frac{C}{G_{S}}\int \text{d}^{3}\text{x}\sqrt{G}\left( \frac{1}{2%
}f\left( \ast T\right) G^{\mu \nu }\partial _{\mu }T\partial _{\nu }T+\cdots
+\theta V\left( \ast T\right) \right) .  \label{maction}
\end{equation}
In this equation $G_{S}$ is the open string coupling constant, $C$ is
related to the $D2l$-brane mass as $C=$\ $G_{S}M_{\text{D2}l}$ and the
effective coupling $f(T)$ is normalised as $f(0)=0$ and $f(t_{\max })=1$ as
suggested by Sen's conjecture.

The total energy $E$ of the scalar field theory eq(\ref{saction}) is given
by 
\begin{equation}
E=\int_{\mathbf{R}^{2}}\text{d}^{2}\text{x}\left[ \left( \partial _{i}\phi
\right) ^{2}+\theta V\left( \ast \phi \right) \right] .  \label{senergy}
\end{equation}
In the large $\theta $ limit, the kinetic term $\left( \partial _{i}\phi
\right) ^{2}$ in eq (\ref{senergy}) may be neglected and the stable field
configuration is achieved by minimising the scalar potential $V(\ast \phi )$%
. Since $V(\ast \phi )$\ is valued in $\mathcal{A}_{\theta }$, its
minimisation is not a simple task as it involves differential Moyal
calculus. A tricky soliton solution has been obtained by Gopakumar, Minwalla
and Strominger (GMS for short); it is based on taking the scalar field $\phi 
$ as \cite{j} 
\begin{eqnarray}
\phi (x) &=&\sum \varphi _{n}p_{n}(x), \\
V(\ast \phi ) &=&\sum V(\varphi _{n})p_{n}(x),
\end{eqnarray}
where the $p_{i}$'s are mutually orthogonal projectors of $\mathcal{A}%
_{\theta }$ and where the $\varphi _{i}$'s are the critical values solving $%
\frac{dV(\varphi )}{d\varphi }=0$. Using Sen's conjecture \ for string field
theory which suggest that the tachyon potential $V(t)$ has two extrema; one
minimum at the origin $t_{\min }=0$ with $V(t_{\min })=0$ and a maximum at $%
t_{\max }$ with $V(t_{\max })$, one looks for special tachyon field
configurations of type $T(x)=t_{\max }p(x)$ solving the equation of motion 
\begin{equation}
\frac{\text{d}V}{\text{d}t}=0,  \label{eqmotion}
\end{equation}
which, upon using the identity$\ p(x)\ast p(x)=p(x),$ should be understood
as 
\begin{equation}
\frac{\text{d}V(tp)}{\text{d}T}|_{T=tp}=\left( \frac{\text{d}V(t)}{\text{d}t}%
\right) p(x).
\end{equation}
The last identity is a special situation of eq(\ref{eqmotion}) and follows
from the fact that any polynomial function $F$ in the projector $p$, we have 
\begin{equation}
F\left( \lambda p\right) =F(\lambda )p.
\end{equation}
Denoting by $a^{-}\equiv a=\frac{1}{\sqrt{2}}\left( x^{1}-ix^{2}\right) $
and $a^{+}=\frac{1}{\sqrt{2}}\left( x^{1}+ix^{2}\right) $, a generic level $k
$ soliton solution is given by projection operators $p_{k}^{(\Lambda )}$,
defined up to unitary $\Lambda $ automorphisms, with $\Lambda \Lambda
^{+}=\Lambda ^{+}\Lambda =I_{d}$ of $\mathcal{A}$ as follows 
\begin{equation}
p_{k}^{(\Lambda )}\left( a^{\pm }\right) =\Lambda ^{+}\left[ \sum_{r=0}^{k-1}%
\frac{1}{r!}\left( a^{+}\right) ^{r}|0\rangle \langle 0|\frac{1}{r!}\left(
a\right) ^{r}\right] \Lambda .  \label{kprojector}
\end{equation}
In this case the soliton solution mimimising (\ref{senergy}) is $\phi
=\varphi _{k}$ $p_{k}^{(\Lambda )}$and the total energy $E_{\min }$ of the
configuration is 
\begin{equation}
E_{\min }=k\theta V(\varphi _{k}).
\end{equation}

Before going ahead, let us give an interpretation of this result in terms of
non-BPS branes and tachyon condensation. For the leading level $k=1,$ eq(\ref
{kprojector}) reduces to $p_{1}(x)=|0\rangle \langle 0|$ or equivalently by
using harmonic oscillator wave functions language 
\begin{equation}
p_{1}\left( x\right) =2\exp \left( -\frac{r^{2}}{\theta }\right) ,
\end{equation}
where we have set $r^{2}=x_{1}^{2}+x_{2}^{2}$. In this case the tachyon
vacuum configuration is 
\begin{equation}
T_{1}(x)=t_{\max }p_{1}(x).
\end{equation}

Note that $t_{\max }$ and $t_{\min }$ have an interesting interpretation;
they describe respectively an unstable local maximum representing the space
filling $D2$-brane, ($V(t_{\max })$), and a local minimum representing the
closed string vacuum without any D-branes ($V(0)=0$). This solution extends
naturally to the $k$-th level as shown here below 
\begin{equation}
T_{k}=t_{\max }\left( |0\rangle \langle 0|+|1\rangle \langle 1|+\ldots
+|k-1\rangle \langle k-1|\right) ;
\end{equation}
it corresponds to $k$ coincident $D0$-branes leading\ then to a $U(k)$ gauge
symmetry.

In what follows we extend the above results for the $D2$-brane on the Moyal
plane to the case of\ a non BPS D$2l$-brane on the $2l$-dimensional Moyal
space introduced earlier and look for new solitonic solutions. Later on, we
shall also consider non-BPS branes on non-commutative torii and explore
their corresponding tachyon solitons.

\subsection{General solitons}

On Moyal space $\mathbf{R}_{\mathbf{\theta }}^{2l}$, the situation is more
general than the previous one and leads to a very rich spectrum containing,
in addition to a stable vaccum field configuration, several other
quasi-stable configurations. As we will see, these configurations constitute
new solitonic solutions not only because they were not considered before but
also because they are associated to$\ $the$\ \left( 2^{l}-2\right) $ local
minima one gets by minimising the total energy of D$2l$-brane in presence of
a NS-NS $B_{IJ}$ field of the form $B_{2i-1,2i}=-B_{2i,2i-1}=B_{i}$ and zero
otherwise. Moreover, these configurations turn out to be very close to
similar ones we will consider later for toric non-BPS D$2l$-branes.

The total energy $E$ of the NC scalar field theory in $\mathbf{R}_{\mathbf{%
\theta }}^{2l}\times \mathbf{R}$ extending eq(\ref{senergy}) associated with
the Moyal plane, reads up to a global normalisation factor as 
\begin{equation}
E=\int_{\mathbf{R}_{\theta }^{2l}}\text{d}^{2l}\text{x}\left\{
\sum_{j=1}^{l}\prod_{i\neq j}\theta _{i}\left[ \left( \partial _{j}\phi
\right) ^{2}+\theta _{j}V\left( \ast \phi \right) \right] \right\} .
\label{gsenergy}
\end{equation}

In eq (\ref{gsenergy}) $V\left( \ast \phi \right) $ is as in (\ref{senergy})
but the star product is now more general; it is deduced from (\ref{star}) by
substituting $\epsilon _{IJ\text{ }}$ by the $2l\times 2l$ antsymmetric
matric $\xi _{2i-1,2i}=1;\quad i=1,\cdots ,l$ and zero otherwise. Moreover,
Weyl correspondence implies that the functions $f$ and $g$ are interpreted
as operators of the the algebra $\mathcal{A}_{\{\theta _{1},\theta
_{2},...,\theta _{l}\}}\mathcal{=}End\mathcal{(H}^{\otimes l}\mathcal{)}$
acting on $\mathcal{H}^{\otimes l}$, the tensorial Hilbert space of $l$
harmonic oscillators. As these $l$ oscillators are uncoupled, the algebra $%
\mathcal{A}_{\{\theta _{1},\theta _{2},...,\theta _{l}\}}\mathcal{=}End%
\mathcal{(H}^{\otimes l}\mathcal{)}$ split as a product of the algebra
factors $\mathcal{A}_{\theta _{i}}$; that is $\mathcal{A}_{\{\theta
_{1},\theta _{2},...,\theta _{l}\}}=\otimes _{i\geq 1}\mathcal{A}_{\theta
_{i}}$. Using this feature, we can rewrite eq (\ref{gsenergy}) in a
remarkable form by introducing the following $E_{j}$ energies 
\begin{equation}
E_{j}=\int_{\mathbf{R}_{\theta }^{2l}}\text{d}^{2l}\text{x}\left\{ \left(
\partial _{j}\phi \right) ^{2}+\theta _{j}V\left( \ast \phi \right) \right\}
,  \label{jgsenrgy}
\end{equation}
in terms of which, the total energy $E$ reads then as 
\begin{equation}
E=\sum_{j=1}^{l}\mu _{j}E_{j},  \label{muenergy}
\end{equation}
where $\mu _{j}$ is given by the product of all $\theta _{i}$'s divided by $%
\theta _{j}$; i.e. 
\begin{equation}
\mu _{j}=\prod_{i\neq j}\theta _{i}=\frac{\left( \prod_{i=1}^{l}\theta
\right) }{\theta _{j}}\equiv \frac{\Theta ^{l}}{\theta _{j}}.
\end{equation}
In the case where all the $\theta _{i}$'s are positive definite, which is
our hypothesis here, the minimum of the total energy $E$ in eq(\ref{muenergy}%
) is achieved by taking the minima of all $E_{i}$'s. These are easily
obtained since for a given fixed $j$, $E_{j}$ is quite similar to the energy
of the NC soliton we have studied for the case of a scalar field theory on $%
\mathbf{R}_{\theta }^{2}\times \mathbf{R}$. Thus taking the large $\theta
_{j}$ limits for all $j$ values, the kinetic energy $\left( \partial
_{j}\phi \right) ^{2}$ of eqs (\ref{gsenergy},\ref{jgsenrgy}) can be
neglegted and the stable configuration is then given by minimising $V(\ast
\phi )$. Extending the GMS formalism to our case where we have $l$ harmonic
oscillators, the soliton solution take the following form 
\begin{equation}
\phi \left( x_{1},\cdots ,x_{2l}\right) =\sum_{n_{1},\cdots ,n_{1}\geq
0}\varphi _{\left( n_{1},\cdots ,n_{l}\right) }P_{\left( n_{1},\cdots
,n_{l}\right) }\left( x_{1},\cdots ,x_{2l}\right) ;
\end{equation}
where now the $P_{\left( n_{1},\cdots ,n_{l}\right) }$'s are mutually
orthogonal projectors of $\mathcal{A}_{\{\theta _{1},\theta _{2},...,\theta
_{l}\}}$ and where the $\varphi _{\left( n_{1},\cdots ,n_{l}\right) }$'s are
as in the Moyal plane analysis. Moreover, introducing the annihilation $\
a_{i}^{-}=a_{i}=\sqrt{\frac{1}{2}}\left( x^{2i-1}-ix^{2i}\right) $ and
creation $a_{i}^{+}=\sqrt{\frac{1}{2}}\left( x^{2i-1}+ix^{2i}\right) $%
operators, one may here also write the $P_{\left( n_{1},\cdots ,n_{l}\right)
}$'s in a form similar to eq(\ref{kprojector}). At a given multi-level$\ 
\mathbf{k}=\left( k_{1},\cdots ,k_{l}\right) ,$ $P_{k}$ is defined up to an
automorphism $\Lambda $ of $\mathcal{A}_{\left\{ \theta _{1},\theta
_{2},...,\theta _{l}\right\} }$, and reads as 
\begin{eqnarray}
P_{k}\left( a_{1}^{\pm },\cdots ,a_{l}^{\pm }\right)  &=&\Lambda
^{+}(\sum_{r_{1}=0}^{k_{1}-1}\sum_{r_{2}=0}^{k_{2}-1}...%
\sum_{r_{l}=0}^{k_{l}-1} \nonumber \\
&&\left\{ \prod_{i}^{l}\left[ \frac{\left( a_{i}^{+}\right) ^{r_{i}}}{\left(
r_{i}\right) !}\right] |0,\cdots ,0\rangle \langle 0,\cdots ,0|\prod_{j}^{l}%
\left[ \frac{\left( a_{i}\right) ^{r_{i}}}{\left( r_{i}\right) !}\right]
\right\} )\Lambda .
\end{eqnarray}
Note that as the $l$ harmonic oscillators are uncoupled, we can simplify the
above relation by using the factorisation property $\mathcal{A}_{\{\theta
_{1},\theta _{2},...,\theta _{l}\}}=\overset{l}{\underset{i=1}{\otimes }}%
\mathcal{A}_{\theta _{i}}$; that is 
\begin{equation}
P_{k}^{\left( \Lambda \right) }=\overset{l}{\underset{i=1}{\otimes }}\left(
P_{ki}^{\left( \Lambda i\right) }\right) ;
\end{equation}
where $P_{ki}^{\left( \Lambda i\right) }$ are projectors in the $i$-th
Hilbert space defined up to $\Lambda _{i}$ automorphisms of $\mathcal{A}%
_{\theta _{i}}$ and read as 
\begin{equation}
P_{ki}^{\left( \Lambda i\right) }=\Lambda _{i}^{+}\left(
\sum_{r_{i}}^{k_{i}-1}\frac{\left( a_{i}^{+}\right) ^{r_{i}}}{\left(
r_{i}\right) !}|0\rangle \langle 0|\frac{\left( a_{i}\right) ^{r_{i}}}{%
\left( r_{i}\right) !}\right) \Lambda _{i}.
\end{equation}
In terms of these projectors, the configuration minimising eq (\ref{gsenergy}%
) is then 
\begin{equation}
\phi \left( a_{1}^{\pm },\cdots ,a_{l}^{\pm }\right) =\varphi _{\mathbf{k}}%
\left[ P_{k_{1}}^{\left( \Lambda _{1}\right) }\left( a_{1}^{\pm }\right)
\otimes P_{k_{2}}^{\left( \Lambda _{2}\right) }\left( a_{2}^{\pm }\right)
\otimes \cdots \otimes P_{k_{l}}^{\left( \Lambda _{l}\right) }\left(
a_{l}^{\pm }\right) \right] ,  \label{fiprojector}
\end{equation}
and the total energy $E^{(0)}$ of the stable solitonic field configuration
to which we shall also refer to as $E_{\text{min}}^{(0)}$ is 
\begin{equation}
E_{\text{min}}^{(0)}=\prod_{i=1}^{l}\left( k_{i}\theta _{i}\right) V\left(
\varphi _{\mathbf{k}}\right) =k\Theta ^{l}V\left( \varphi _{\mathbf{k}%
}\right) ,  \label{abminima}
\end{equation}
where we have also set $k=\prod_{i=1}^{l}k_{i}$. Up to now this analysis
seems to be a generalisation of the analysis performed for the Moyal plane.
However this is not the full story since the minimisation of the total
energy eq (\ref{gsenergy}) leads also to other local minima depending on the
various ways large non-commutativity is taken. In what follows, we study
these local minima and explore the field configurations associated with them
as well as their interpretation in terms of non-BPS D-branes.

To start reconsider eq (\ref{gsenergy}) and reexamine all its possible local
minima. Since all the coefficients $\mu _{j}$ in front of $E_{j}$ are
positive, then the total energy minimum $E_{\text{min}}$ obtained by taking
the minimum of all $E_{j}$'s. In other words $E_{\text{min}}$, which in
addition to the kinetic and potential energies depends moreover on the
magnitudes of the $\theta _{i}$'s parametes, is given by 
\begin{equation}
E_{\text{min}}(\theta _{1,}\theta _{2},...,\theta _{l})=\sum_{j=1}^{l}\mu
_{j}(E_{j})_{\text{min}}.  \label{minenergy}
\end{equation}

From this equation, one clearly see that the value of $E_{\min }$ depends on
the ways the large non-commutativity limit is taken. If one adopts a strong
definition of large non-commutativity by requiring all $\theta _{i}$'s
large, then all ($E_{j})_{\min }$'s are equal to $k_{j}\theta _{j}V(\varphi
_{k})$ and so one discovers the absolute minimum $E_{\min }^{(0)}$ given by
eq(\ref{abminima}). However, if one adopts a weaker definition for large
non-commutativity by requiring that at least one of the $\theta _{i}$'s\ is
large, then the above equation will have several local minima. Let us
determine with explicit details the two leading ones and give the general
result using iteration techniques.

\bigskip (1) $E_{\text{min}}^{(1)}$\textbf{\ local minima energy}

This energy is obtained from eq(\ref{minenergy}) by taking all $\theta _{i}$%
's large except one of them, say $\theta _{j}$ for some given $j$, which
taken finite. Since $j$ can take $l$ values, the $E_{\text{min}}^{(1)}$
energy is $l$-th degenerate. Indeed, within this limit one has $(E_{i})_{%
\text{min},i\neq j}=k_{i}\theta _{i}V(\varphi _{k})$ in agreement with eqs (%
\ref{gsenergy},\ref{jgsenrgy}) and can usually set the $(E_{j})_{\text{min}}$
energy associated to the finite $\theta _{j}$ as given by a derivation above 
$k_{j}\theta _{j}V(\varphi _{k})$. Setting$\ k_{n}\theta _{n}V(\varphi
_{k})=E_{n}^{0}$\ and $(E_{j})_{\text{min}}=E_{j}^{0}+\delta E_{j}^{0}$,
where $\delta E_{j}^{0}$ is the gap energy, then putting back into eq(\ref
{minenergy}), one gets 
\begin{equation}
(E_{\text{min}}^{(1)})_{j}=E_{\text{min}}^{(0)}+\delta E_{j}^{0};\quad
j=1,...,l.  \label{jminenrgy}
\end{equation}
Degeneracy of ($E_{\text{min}}^{(1)})_{j}$ is ensured if one assumes that
all $\delta E_{j}^{0}$ are equal otherwise the degeneracy is rised either
partially or totaly depending on the values of$\ \delta E_{j}^{0}$ and so
one ends with different quasi-stable field configurations.

\bigskip (2) $E_{\text{min}}^{(2)}$\textbf{\ local minima energy}

In tis case the energy $E_{\text{min}}^{(2)}$ is obtained from eq(\ref
{minenergy}) by taking all $\theta _{i}$'s large except two of them, say $%
\theta _{m}$ and $\theta _{n}$ which are taken to be finite. Similar
analysis as before shows that $E_{\text{min}}^{(2)}$ depends on two indices $%
m$ and $n$ and reads as 
\begin{equation}
(E_{\text{min}}^{(2)})_{\{m,n\}}=E_{\text{min}}^{(0)}+\delta
E_{m}^{0}+\delta E_{n}^{0};\quad m,n=1,...,l.
\end{equation}
If we assume that $\delta E_{m}^{0}$ and $\delta E_{n}^{0}$ are equal, then
the corresponding configuration is $\frac{l(l-1)}{2}$degenerate.

More generally if we assume that large non-commutativity is achieved by
taking the set of parameters $\left\{ \theta _{i_{1}},\theta
_{i_{2}},...,\theta _{i_{s}}\right\} $ finite\ and $\left\{ \theta
_{i_{s+1}},\theta _{i_{s+2}},...,\theta _{i_{l}}\right\} $ large. The energy
of the local minimum is $(E_{\text{min}}^{(s)})_{\{i_{1},i_{2},...,i_{s}\}}$%
\ reads as 
\begin{equation}
(E_{\text{min}}^{(s)})_{\{i_{1},i_{2},...,i_{s}\}}=E_{\text{min}%
}^{(0)}+\sum_{1\leq j\leq s}\delta E_{i_{j}}^{0}\quad ;i_{j}=1,...,l;0\leq
s\leq l-1.
\end{equation}

(3) $E_{\min }^{(s)}$\textbf{\ local minima energy; }$3\leq s\leq l-1.$

Here we give a recaputilating table (a) where we put the energies of local
minima, their maximal degeneracies as well as a potential curve $V(\phi )$
representing these minima

\begin{center}
\begin{tabular}{|c|c|}
\hline
Minima energies & \# of degenerate states \\ \hline
$E_{\text{min}}^{\left( 0\right) }$ & 1 \\ \hline
$E_{\text{min}}^{\left( 1\right) }$ & $l$ \\ \hline
$E_{\text{min}}^{\left( 2\right) },....$ & $\frac{l\left( l-1\right) }{2},...
$ \\ \hline
$E_{\text{min}}^{\left( s\right) }$ & $\frac{l!}{\left( l-s\right) !s!}$ \\ 
\hline
$E_{\text{min}}^{\left( l-1\right) }$ & $l$ \\ \hline
\end{tabular}
\end{center}

\bigskip 

\begin{center}
Table (a)
\end{center}

\begin{figure}[tbh]
\begin{center}
\epsfig{file=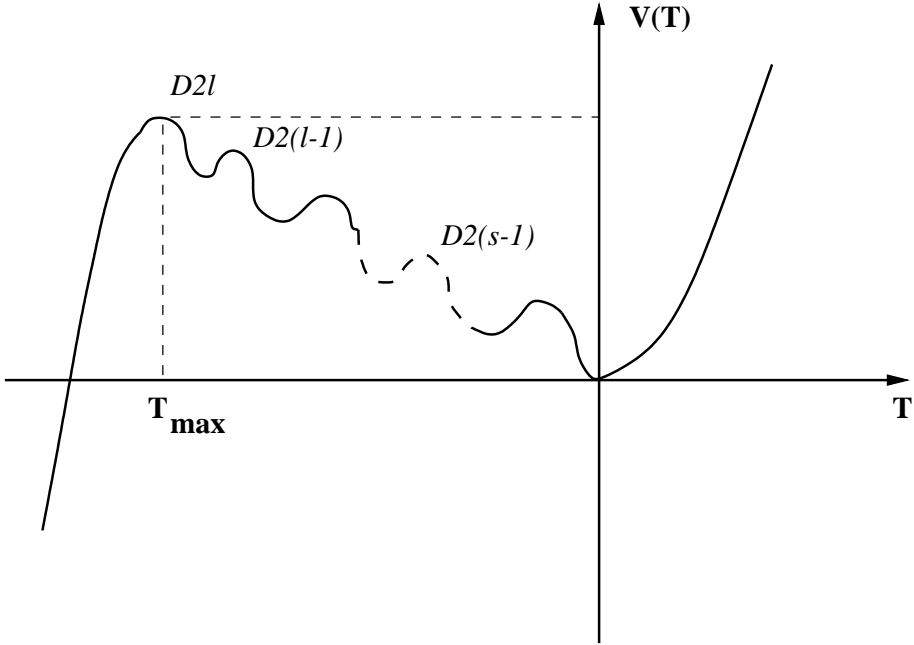}
\end{center}
\caption{the suggested shape of the potentil tachyon describing the decay of
D$2l$-brane. To each local maximum $s$ is associated a non-BPS D$2s$-brane.}
\label{fig:cur}
\end{figure}

The total number of local minima $(E_{\text{min}}^{(s)})_{%
\{i_{1},i_{2},...,i_{s}\}}$, $\left( 1\leq s\leq l-1\right) $, is then $%
\left( 2^{l}-2\right) $ to which one should add the absolute minimum $E_{%
\text{min}}^{\left( 0\right) }$ and the upper bound where all $\theta _{i}$%
's are taken finite. On the figure \ref{fig:cur} representing a potential
curve, which may be though of as the potential of the non-commutative scalar
field in Moyal space, we have represented the various local local minima. 
In string field theory these local minima might be interpreted as associated
with non-BPS states; the recipe is that to the absolute minimum we encounter 
$\prod_{i=1}^{l}(k_{i})$ non-BPS D$0$-brane with a $U(\prod_{i=1}^{l}(k_{i}))
$ gauge symmetry while for the leading local minima of energies $(E_{\text{%
min}}^{(1)})_{j}$, we have, for fixed $j$, $\prod_{i=1,i\neq
j}^{l}(k_{i})=\prod_{i=1}^{l}(k_{i})/k_{j}$ non-BPS D$2$-branes with a $%
U(\prod_{i=1}^{l}(k_{i})/k_{j})$ gauge symmetry and whose world volume
includes the $(x^{2j-1},x^{2j})$ Moyal plane itself contained in the $2l$%
-dimensional Moyal space. Therefore, the total gauge group of non-BPS\ D$2$%
-branes is the cross product of $U(\prod_{i=1}^{l}(k_{i})/k_{j})$, that is $%
\otimes _{j=1}^{l}U(\prod_{i=1}^{l}(k_{i})/k_{j})$. More generally to the
local minima of energies $(E_{\text{min}}^{(s)})_{\{i_{1},i_{2},...,i_{s}\}}$%
, $\left( 0\leq s\leq l\right) $, we have for a given configuration $%
(\prod_{i=1}^{l}k_{i})/(\prod_{n=1}^{s}k_{i_{n}})\ $non-BPS D$2s$-branes
with a\ $U\left( (\prod_{i=1}^{l}k_{i})/(\prod_{n=1}^{s}k_{i_{n}})\right) $
gauge symmetry. The total gauge group $G_{s}$ may here also be written down;
it is given by the tensor product of the $\frac{l!}{\left( l-s\right) !s!}$
possible factors. It reads as 
\begin{equation}
G_{s}=\otimes _{j=1}^{s}\otimes _{i_{j}=1}^{l}U\left( \frac{%
\prod_{i=1}^{l}(k_{i})}{\prod_{n=1}^{s}k_{i_{n}}}\right) .
\end{equation}

Now we turn to give the corresponding field configurations associated to
these local minima. These solutions may be written down explicitly\ by using
the appropriate projector opertors on these vacuua. To see how these are
built let us first study the leading examples, then extends the results to
all local minima. For the absolute minimum $E_{\text{min}}^{(0)}=\left(
\prod_{i=1}^{l}k_{i}\right) \Theta ^{l}V\left( \varphi \right) $, the
corresponding soliton denoted as $\phi ^{(0)}=\phi _{j}^{\left( 0\right)
}(\{a_{i}^{\pm }\})$, was already calculated as shown in eq(\ref{fiprojector}%
); it reads as 
\begin{equation}
\phi ^{(0)}=\varphi P_{k_{1}}\otimes P_{k_{2}}\otimes \ldots \otimes
P_{k_{l}},
\end{equation}
where we have dropped the $\Lambda _{i}$ indices on the $P_{k_{i}}$'s for
simplicity. This is also equivalent to choose $\Lambda _{i}$'s as the
identity operator This configuration is a stable non-degenerate state and
may be thought of as the closed string vacuum in tachyon condensation
framework. For the\ field configurations associated to the local energy
minima $E_{j}^{\left( 1\right) }$ eq(\ref{jminenrgy}), one has $l$
quasi-stable states which we denote as $\phi _{j}^{\left( 1\right) }=\phi
_{j}^{\left( 1\right) }(\{a_{i}^{\pm }\}_{i\neq j})$. They are are given by 
\begin{eqnarray}
\phi _{1}^{\left( 1\right) } &\backsim &P_{1}^{\left( 1\right) }=\mathbf{1}%
\otimes P_{k_{2}}\otimes \ldots \otimes P_{k_{l}} \\
&&\vdots   \notag \\
\phi _{n}^{\left( 1\right) } &\backsim &P_{n}^{\left( 1\right)
}=P_{k_{1}}\otimes P_{k_{2}}\otimes \ldots \otimes \mathbf{1}_{n}\otimes
\ldots \otimes P_{k_{l}} \\
&&\vdots   \notag \\
\phi _{l}^{\left( 1\right) } &\backsim &P_{l}^{\left( 1\right)
}=P_{k_{1}}\otimes P_{k_{2}}\otimes \ldots \otimes P_{k_{l-1}}\otimes 
\mathbf{1,}
\end{eqnarray}
where $P_{j}^{\left( 1\right) }$ stands for the projector operator the first
local minimum with an identity operator at the $j$-th position. Similarly,
we have $l\left( l-1\right) /2$ quasi-stable states $\phi
_{\{j_{1},j_{2}\}}^{\left( 2\right) }=\phi _{\{j_{1},j_{2}\}}^{\left(
2\right) }(\{a_{i}^{\pm }\}_{i\neq \{j_{1},j_{2}\}})$ associated with the
local minima energy $E_{\text{min}}^{\left( 2\right) }$; they read as 
\begin{eqnarray}
\phi _{\{1,2\}}^{\left( 2\right) } &\backsim &P_{\{1,2\}}^{\left( 2\right) }=%
\mathbf{1}\otimes \mathbf{1}\otimes P_{k_{3}}\otimes \ldots \otimes P_{k_{l}}
\\
&&\vdots   \notag \\
\phi _{\{j_{1},j_{2}\}}^{\left( 2\right) } &\backsim
&P_{\{j_{1},j_{2}\}}^{\left( 2\right) }=P_{k_{1}}\otimes P_{k_{2}}\otimes
\ldots \otimes \mathbf{1}_{j_{1}}\otimes \ldots \otimes \mathbf{1}%
_{j_{2}}\otimes \ldots \otimes P_{k_{l}} \\
&&\vdots   \notag \\
\phi _{\{l-1_{1},l\}}^{\left( 2\right) } &\backsim
&P_{\{l-1_{1},l\}}^{\left( 2\right) }=P_{k_{1}}\otimes P_{k_{2}}\otimes
\ldots \otimes P_{k_{l-1}}\otimes \mathbf{1}_{l-1}\mathbf{\otimes \mathbf{1}}%
_{l}\mathbf{.}
\end{eqnarray}
More generally, the vacuum field configurations $\phi
_{\{j_{1},j_{2},...,j_{s}\}}^{\left( s\right) }=\phi
_{\{j_{1},j_{2},...,j_{s}\}}^{\left( s\right) }\{a_{i}^{\pm }\})$, for $%
i\neq \{j_{1},j_{2},...,j_{s}\}$ and $0\leq s\leq l-1$, associated with a
generic local minima of energy $(E_{\min }^{(s)})_{\{j_{1},j_{2},...,j_{s}\}}
$ are given by 
\begin{equation}
\phi _{\{j_{1},j_{2},...,j_{s}\}}^{\left( s\right) }\backsim
P_{\{j_{1},j_{2},...,j_{s}\}}^{\left( s\right) }=P_{k_{1}}\otimes \ldots
\otimes \mathbf{1}_{j_{1}}\otimes \ldots \otimes P_{k_{n}}\otimes ...\otimes 
\mathbf{1}_{j_{s}}\otimes \ldots \otimes P_{k_{l}}.
\end{equation}

The soliton solutions we have been desribed possede an interesting
application in string field theory \cite{k} where the vacuum scalar field
configurations are identified with tachyon solitons of various non-BPS
D-branes.

To conclude this section we should say that\ all the analysis concerning the
NC soliton of D$2$-brane on the Moyal plane and which was interpreted as
describing D$0$-branes which appear after tachyon condensation giving the
correct D$0$-brane mass as well as their number in the vacuum state extends
naturally to the case of D$2l$-branes on Moyal spaces. The novelty here is
that one has in addition to the D$0$-branes and D$2l$ ones other
quasi-stable configurations which we have interpreted as extra D$2s$-branes; 
$1\leq s\leq l-1$, living altogether with D$0$ and D$2l$ ones.\ Moreover
given a D$2s$-brane, one distinguishes different world volumes for these
branes; $l$ kinds of D$2$-branes, $l(l-1)/2$ kinds of D$4$-branes and so on.
Decomposing the $P_{\{j_{1},j_{2},...,j_{s}\}}^{\left( s\right) }$
projectors into mutually irreducible orthogonal $\chi
_{\{r_{1},r_{2},...,r_{s}\}}^{(s)}$ ones as herebelow 
\begin{equation}
P_{\{j_{1},j_{2},...,j_{s}\}}^{\left( s\right)
}=\sum_{r_{1}=0}^{j_{1}-1}\sum_{r_{2}=0}^{j_{2}-1}...%
\sum_{r_{l}=0}^{j_{l}-1}\chi _{\{r_{1},r_{2},...,r_{s}\}}^{(s)},
\end{equation}
with 
\begin{eqnarray}
(\chi _{\{r_{1},r_{2},...,r_{s}\}}^{(s)})^{2} &=&\chi
_{\{r_{1},r_{2},...,r_{s}\}}^{(s)};\quad (\chi
_{\{r_{1},r_{2},...,r_{s}\}}^{(s)})^{\dagger }=\chi
_{\{r_{1},r_{2},...,r_{s}\}}^{(s)} \\
\chi _{\{r_{1},r_{2},...,r_{s_{1}}\}}^{(s_{1})}\chi
_{\{q_{1},q_{2},...,q_{s_{2}}\}}^{(s_{2})} &=&0;\quad \qquad \{_{s_{1}=s_{2}%
\text{ with }\{r_{1},r_{2},...,r_{s_{1}}\}\neq
\{q_{1},q_{2},...,q_{s_{2}}\}}^{s_{1}\neq s_{2}\text{ or }}
\end{eqnarray}
or equivalently by rewriting them in a short form as 
\begin{equation}
P_{\mathbf{j}}^{\left( s\right) }=\sum_{\mathbf{r}}^{\mathbf{j}-\mathbf{1}%
}\chi _{\mathbf{r}}^{(s)},
\end{equation}
and 
\begin{eqnarray}
(\chi _{\mathbf{r}}^{(s)})^{2} &=&\chi _{\mathbf{r}}^{(s)};\quad (\chi _{%
\mathbf{r}}^{(s)})^{\dagger }=\chi _{\mathbf{r}}^{(s)}  \notag \\
\chi _{\mathbf{r}_{1}}^{(s_{1})}\chi _{\mathbf{r}_{2}}^{(s_{2})} &=&0;\quad
for\text{ }s_{1}\neq s_{2}\text{ or }s_{1}=s_{2}\text{ but }\mathbf{r}%
_{1}\neq \mathbf{r}_{2},
\end{eqnarray}
then the open string wave function $\Psi $ can be projected into pieces $%
\chi _{\mathbf{r}}^{(s_{1})}\Psi \chi _{\mathbf{q}}^{(s_{2})}$ representing
the open strings interpolating from a non-BPS D$2s_{1}$-brane with data $%
\mathbf{r}=(r_{1},r_{2},...,r_{s_{1}})$ to the non BPS D$2s_{2}$ one with
data $\mathbf{q}=(q_{1},q_{2},...,q_{s_{1}})$. Therefore one has a variety
of open strings ending on D-branes; in particular D$0$-D$0$, D$0$-D$2$, D$0$%
-D$4$,$\cdots $, D$2$-D$2$, D$2$-D$4$,$\cdots $; D$4$-D$4$,$\cdots $ and so
on.

\section{Solitons in non-commutative torus $\mathbb{T}_{\protect\theta
}^{2l} $}

Here we want to extend the analysis we have made for Moyal space to the
non-commutative torus $\mathbb{T}_{\mathbf{\theta }}^{2l}$. Like for eqs (%
\ref{ncomm}), the $\mathbb{T}_{\mathbf{\theta }}^{2l}$ we will be
considering is roughly speaking given by the product of $l$ non-commutative
two dimensional torii $\mathbb{T}_{\theta _{i}}^{2}$. In other words the
non-commutative $\mathbb{T}_{\mathbf{\theta }}^{2l}$ is generated by a
system of $l$ unitary pairs $\left( U_{i},V_{i}\right) $ satisfying the
algebra 
\begin{eqnarray}
U_{n}V_{n} &=&\text{e}^{-i2\pi \theta _{n}}V_{n}U_{n}  \label{nctorus} \\
U_{n}V_{m} &=&V_{m}U_{n};\quad n\neq m,  \label{torus}
\end{eqnarray}

for which we will shall refer from now on as $\mathfrak{A}_{\theta }$. Note that
because of eq(\ref{torus}), the non-commutative algebra $\mathfrak{A}_{\theta }$
may be also defined as the tensor product of $l$\ factors $\mathfrak{A}_{\theta
_{i}}$; $\mathfrak{A}_{\theta }=\otimes _{i=1}^{l}\mathfrak{A}_{\theta _{i}}$; Each $%
\mathfrak{A}_{\theta _{i}}$ factor is associated with the non-commutative torus $%
\mathbb{T}_{\theta _{i}}^{2}$; and the corresponding $\left(
U_{i},V_{i}\right) $ pairs are realised as the exponentials of the
non-commutative coordinates $\left( x^{2i-1},x^{2i}\right) $ of $\mathbb{T}%
_{\theta _{i}}^{2}$. For later use we prefer to denote the coordinates of
the non-commutative torus by the capital letters $\left(
X^{2i-1},X^{2i}\right) $ while those of the commutative ones by small
letters. Thus we have 
\begin{equation}
U_{i}=\text{e}^{\frac{i2\pi }{R_{2i-1}}X^{2i-1}};\quad V_{i}=\text{e}^{\frac{%
i2\pi }{R_{2i}}X^{2i}},\quad i=1,\ldots ,l  \label{geomtorus}
\end{equation}
where the $R_{j}$'s are the one cycles radii of the $2l$-dimensional torus.
Note also that $U_{i}$ and $V_{i}$ generators have different representations
according to whether the $\theta _{i}$'s are rational or irrational. In what
follows we shall use both of these representations; this is why we first review
them briefly on the simple case of $%
\mathbb{T}_{\theta _{i}}^{2}$ torus.

\bigskip

\textit{Rational representations }

\quad This kind of representations\ corresponds to rational values of $%
\theta _{i}=q_{i}/p_{i}$, where $p_{i}$ and $q_{i}$ are\ mutually coprime
integers. The $U_{i}$ and $V_{i}$ generators are given by the following
finite $p_{i}\times p_{i}$ matrices 
\begin{equation}
U_{i}=\left[ 
\begin{array}{ccccc}
1 & 0 & 0 & \cdots  & 0 \\ 
0 & \omega _{i} & 0 & \cdots  & 0 \\ 
0 & 0 & \omega _{i}^{2} & \cdots  & 0 \\ 
\cdots  & \cdots  & \cdots  & \cdots  & 0 \\ 
0 & 0 & 0 & \cdots  & \omega _{i}^{p_{i}-1}
\end{array}
\right] ;\quad \quad V_{i}=\left[ 
\begin{array}{ccccc}
0 & 1 & 0 & \cdots  & 0 \\ 
0 & 0 & 1 & \cdots  & 0 \\ 
0 & 0 & 0 & \cdots  & 0 \\ 
\cdots  & \cdots  & \cdots  & \cdots  & 1 \\ 
1 & 0 & 0 & \cdots  & 0
\end{array}
\right]   \label{matrices}
\end{equation}
where $\omega _{i}=$e$^{i2\pi q_{i}/p_{i}}$. Note in passing that $%
U_{i}^{p_{i}}$ and $V_{i}^{p_{i}}$ act as the $p_{i}\times p_{i}$ identity
operator $I$ and so any element $a_{i}$ of the non-commutative algebra $%
\mathfrak{A}_{\theta _{i}}$ associated to $\mathbb{T}_{\theta _{i}}^{2}$ has a
finite expansion 
\begin{equation}
a_{i}=\sum_{n,m=0}^{p-1}(a_{i})_{nm}U_{i}^{n}V_{i}^{m}.
\end{equation}

Note that in the matrix representation presented above, the $U_{i}$
generator is given by a diagonal matrix; a feature which allows to build the
usual rank $k_{i}$ projector ($\Pi _{i})_{k_{i}}=$diag$\left( 1,1,\ldots
,1,0,\ldots ,0\right) $ as a series of the $U_{i}$'s, i.e. 
\begin{equation}
(\Pi _{i})_{k_{i}}=\sum_{n=0}^{p}(a_{i})_{n0}U_{i}^{n}.  \label{repprojector}
\end{equation}

A direct check shows that the ($a_{i})_{n0}$ coefficients are given by ($%
a_{i})_{n0}=\frac{1}{p_{i}}\frac{1-\omega _{i}^{-nk}}{1-\omega _{i}^{-n}}$.
Note moreover that the trace on $\mathfrak{A}_{\theta _{i}}$\ is defined as Tr($%
\Pi _{i})_{k_{i}}=(a_{i})_{00}=k_{i}$. Thus the range of $k_{i}$\ is $0\leq
k_{i}\leq p_{i}$ and is interpreted as the number of D0-branes one obtains
from the study of a D$2$-brane on the non-commutative $\mathbb{T}_{\theta
_{i}}^{2}$.

\bigskip

\textit{Irrational representations}

\quad The generalisation of the previous case to irrational $\theta _{i}$'s
is not automatic and turns out to have interesting interpretations in terms
of branes bound states. Following the same lines as for the rational case by
working in a representation in which $U_{i}$ is diagonal and $V_{i}$ is not,
one has the following 
\begin{eqnarray}
\langle x_{2i-1}^{\prime }|U_{i}|x_{2i-1}\rangle  &=&\text{e}^{i2\pi
x_{1}}\delta \left( x_{2i-1}^{\prime }-x_{2i-1}\right) ;  \notag \\
\langle x_{2i-1}^{\prime }|V_{i}|x_{2i-1}\rangle  &=&\delta \left(
x_{2i-1}+\theta _{i}-x_{2i-1}^{\prime }\right) ,  \label{irrrep}
\end{eqnarray}
where we have set $R_{I}=1$ for commodity. Note in passing that here also $%
U_{i}$ is a diagonal operator while\ $V_{i}$\ is not; they depend on the $%
x^{1}$ variable only. To construct the projector operators on the position
space generated by the continuous basis vectors $\left\{ |x_{2i-1}\rangle
\times |x_{2i}\rangle \right\} $, one may consider in a first attempt
functions of the diagonal operator $U_{i}$. A choice of the function $%
f(U_{i})$ is given by 
\begin{eqnarray}
\langle x_{2i-1}^{\prime }|(\Pi _{i})|x_{2i-1}\rangle  &=&\langle
x_{2i-1}^{\prime }|f(U_{i})|x_{2i-1}\rangle   \notag \\
&=&\theta ^{IJ}=\left\{ 
\begin{array}{ll}
\kappa _{i}\delta \left( x_{2i-1}^{\prime }-x_{2i-1}\right) , & 0\leq
x_{2i-1}\leq \kappa _{i} \\ 
0 & \kappa _{i}<x_{2i-1}\leq 1
\end{array}
\right. 
\end{eqnarray}
$\kappa _{i}$ is a priori a real parameter lying between zero and one. This
choice of ($\Pi _{i})$\ ensures that it is hermitian, ($\Pi _{i})^{2}=(\Pi
_{i})$ but still fails as,in general, the trace Tr($\Pi _{i})$ is not an
integer
\begin{equation}
\text{Tr(}\Pi _{i})=\int \text{d}x_{2i-1}\langle x_{2i-1}|(\Pi
_{i})|x_{2i-1}\rangle =\kappa _{i}.
\end{equation}
This trace is not acceptable, it contradicts the expected spectrum dictated
by the group $\mathbf{K}_{0}\left( \mathfrak{A}_{\theta _{i}}\right) =\mathbb{Z}%
+\theta _{i}\mathbb{Z}$, since $\kappa _{i}$ is not quantised. To overcome
this difficulty one should use both the $U_{i}$ and $V_{i}$ operators
instead of using $U_{i}$ alone; this will allow to also incorporate the
non-commutativity in the game. A class of solutions for the projector
operators in agreement with $\mathbf{K}_{0}\left( \mathfrak{A}_{\theta
_{i}}\right) $ has been constructed in \cite{p}. It extends the
Powers-Rieffel projectors and reads as 
\begin{equation}
\mathcal{P}_{\left\{ n_{i}+m_{i}\theta _{i}\right\} }=\left(
V_{i}^{m_{i}}\right) ^{+}\left( g\left( U_{i}\right) \right) ^{+}+f\left(
U_{i}\right) +g\left( U_{i}\right) V_{i}^{m_{i}};  \label{irrprojector}
\end{equation}
where the function $f(U_{i})$ and $g(U_{i})$ are given by 
\begin{equation}
f\left( U_{i}\right) =\left\{ 
\begin{array}{ll}
x^{2i-1}/\epsilon _{i} & x^{2i-1}\in \left[ 0,\epsilon _{i}\right]  \\ 
1 & x^{2i-1}\in \left[ \epsilon _{i},\theta _{i}\right]  \\ 
1-\left( x^{2i-1}-(n_i+m_i \theta _{i})\right) /\epsilon _{i} & x^{2i-1}\in \left[
\theta _{i},\theta _{i}+\epsilon _{i}\right]  \\ 
0 & x^{2i-1}\in \left[ \theta _{i}+\epsilon _{i},1\right] 
\end{array}
\right. 
\end{equation}
\begin{equation}
g\left( U_{i}\right) =\left\{ 
\begin{array}{ll}
\sqrt{f\left( U_{i}\right) \left( 1-f\left( U_{i}\right) \right) } & 
x^{2i-1}\in \left[ 0,\epsilon _{i}\right]  \\ 
0 & x^{2i-1}\in \left[ \epsilon _{i},1\right] 
\end{array}
\right. .
\end{equation}
In these eqs\ $\epsilon _{i}$\ is a small parameter which physically may be
interpreted as a regulation parameter. Having given the representations of $%
\mathfrak{A}_{\theta _{i}}$ for $\mathbb{T}_{\theta _{i}}^{2}$, we turn now to
extend them to $\mathbb{T}_{\mathbf{\theta }}^{2l}$. For fixed $l$, we have
generally $2^{l}$ possibilities depending on whether the $\theta _{i}$'s are
rational or irrational. If all $\theta _{i}$'s are rational, i.e. $\theta
_{i}=q_{i}/p_{i}$ the $U_{i}$ and $V_{i}$ are given by similar eqs to eq (%
\ref{matrices}). If instead all $\theta _{i}$'s are irrational, the $U_{i}$%
's and $V_{i}$'s are given by 
\begin{eqnarray}
\langle \mathbf{x}^{\prime }|U_{i}|\mathbf{x}\rangle  &=&\text{e}^{i2\pi
x_{i}}\delta \left( \mathbf{x}^{\prime }-\mathbf{x}\right) , \\
\langle \mathbf{x}^{\prime }|V_{i}|\mathbf{x}\rangle  &=&\delta ^{2l}\left( 
\mathbf{x}+\theta _{i}-\mathbf{x}^{\prime }\right) .
\end{eqnarray}

We can also have the case where part of the $\theta _{i}$'s are rational and
the others are irrational. In this case the $U_{i}$' s and $V_{i}$'s are
given by mixing the representations (\ref{matrices}) and (\ref{irrrep}).

The projectors $\mathcal{P}_{\left\{ \theta _{1},\ldots ,\theta _{l}\right\}
}$ on the space position basis $\{|\mathbf{x}\rangle
=|(x_{1},x_{2},...,x_{2l-1},x_{2l})\rangle \}$ for $\mathbb{T}_{\mathbf{%
\theta }}^{2l}$ have then several forms depending on whether the $\theta _{i}
$'s are rational or irrational. Denoting by $\mathcal{P}_{\left\{ \theta
_{i}\right\} }$ the projector operator associated to $\theta _{i}$ which is
given by either eq (\ref{repprojector}) or (\ref{irrprojector}), we have 
\begin{equation}
\mathcal{P}_{\left\{ \theta _{1},\ldots ,\theta _{l}\right\} }=\overset{l}{%
\underset{i=1}{\otimes }}\mathcal{P}_{\left\{ \theta _{i}\right\} }.
\label{gtprojector}
\end{equation}

From eq (\ref{gtprojector}), one learns that there are a priori $2^{l}$
solutions. However if one identifies operators that are related under
permutations of positions, one ends then with $l$ differents objects. Note
that the trace of eq (\ref{gtprojector}) is given by the trace on the
individual projectors $\mathcal{P}_{\left\{ \theta _{i}\right\} }$; i.e 
\begin{equation}
\text{Tr}\mathcal{P}_{\left\{ \theta _{1},\ldots ,\theta _{l}\right\}
}=\prod_{i=1}^{l}\text{Tr}\mathcal{P}_{\left\{ \theta _{i}\right\} }.
\label{traceproj}
\end{equation}

\section{Non-BPS Toric D$2l-$Branes}

Consider a non-BPS D$2l$-brane on the $2l$ dimensional non-commutative torus 
$\mathbb{T}_{\mathbf{\theta }}^{2l}$ given by eq(\ref{geomtorus}) and study
the field configurations minimising the total energy $E(T)$ of the tachyon
living on the world volume of the brane. Starting from the D$2l$-brane of
string theory ($0\leq l\leq 12$ for the bosonoic case and $0\leq l\leq 4$
for type IIB) in presence of $B_{\mu \nu }$ field chosen as in the case of
the Moyal space, the string field theory effective action $\mathcal{S}=%
\mathcal{S}(T(x))$, keeping only the tachyon field $T(x)$ and integrating
out all other fields, reads as 
\begin{equation}
\mathcal{S}=\frac{C_{\text{D}2l}}{G_{S}}\int \text{d}^{2l+1}\text{x}\sqrt{G}%
\left( \frac{1}{2}f\left( \ast T\right) G^{\mu \nu }\partial _{\mu
}T\partial _{\nu }T+\cdots +V\left( \ast T\right) \right) ,  \label{taction}
\end{equation}
where $G_{\mu \nu }$, $G_{S}$, $C_{\text{D}2l}$ and the factor $f\left(
t\right) $ are as in eq(\ref{maction}).

In large non-commutativity, the kinetic term of the tachyon is neglected and
so the total energy $E(T)$\ reduces to 
\begin{equation}
E(T)=M_{\text{D2}l}\text{Tr }V(T),
\end{equation}
where $M_{\text{D2}l}$ denotes the mass of the original D2$l$-brane and the
trace is normalysed as Tr $\mathbf{1}=1$. Extremisation of $E(T)$\ is
achieved as usual by using the GMS approach\ which shows that tachyon field
configuration are proportional to projectors in the $\mathfrak{A}_{\mathbf{%
\theta }}$ non-commutative algebra.\ The interpretation of the solution
field configurations are given by Sen's conjecture, where the original D2$l$%
-brane is interpreted as to correspond to $T=t_{\text{max}}\mathbf{1}$ and
the complete tachyon condensation $\left( T=0\right) $ to the decay at the
vacuum.

In the case of non-BPS D$2l$-brane on the $2l$ dimensional non-commutative
torus $\mathbb{T}_{\mathbf{\theta }}^{2l}$ we are interested in here, the
tachyon field configurations extremising eq (\ref{taction}) is given by $%
T=t_{\text{max}}\mathcal{P}_{\left\{ \theta _{1},\ldots ,\theta _{l}\right\}
}$. Using eq (\ref{traceproj}) and taking all $\theta _{i}$'s irrational,
the total energy of the soliton is 
\begin{equation}
E_{\text{total}}\equiv E\left( \mathcal{P}_{\left\{ n_{1}+m_{1}\theta
_{1},\ldots ,n_{l}+m_{l}\theta _{l}\right\} }\right) =M_{\text{D2}%
l}\prod_{i=1}^{l}\left( n_{i}+m_{i}\theta _{i}\right) .  \label{totalmass}
\end{equation}
Taking into account the fact that$\ 0\leq n_{i}+m_{i}\theta _{i}\leq 1$ and
so their product, one sees that the energy (\ref{totalmass}) is bounded by
the mass of the original D2$l$-brane 
\begin{equation}
E_{\text{total}}\leq M_{\text{D2}l}.
\end{equation}
Moreover expanding eq (\ref{totalmass}) in $\theta _{i}$ series as 
\begin{equation}
E_{\text{total}}=N+\sum_{i=1}^{l}\theta _{i}N_{i}+\sum \theta
_{ij}N_{ij}+\cdots +\theta _{i_{1}\cdots i_{l}}N_{i_{1}\cdots i_{l}}
\label{energydevlop}
\end{equation}
with 
\begin{eqnarray}
N &=&\prod_{i}n_{i},...\theta _{i_{1}\cdots i_{s}}=\prod_{i=1}^{s}\theta
_{i};\quad N_{i}=\frac{m_{i}}{n_{i}}N;\quad N_{ij}=\frac{m_{j}}{n_{j}}%
N_{i};...  \notag \\
N_{i_{1}i_{2}}.._{.i_{s}} &=&\frac{m_{i_{s}}}{n_{i_{s}}}%
N_{i_{1}i_{2}}...i_{s-1},\left( 1\leq s\leq l-1\right) ;\quad N_{i_{1}\cdots
i_{l}}=\prod_{i}m_{i}.
\end{eqnarray}
We will turn in a moment to this expansion, but now let us the tools we will
need by giving the relations between the masses of D2$j$-branes $1\leq j\leq
l$. For a generic $j$, the masses of non-BPS branes on a non-commutative $%
\mathbb{T}_{\mathbf{\theta }}^{2j}$ torii read as 
\begin{equation}
M_{\text{D}2j}=\sqrt{2}\frac{\prod_{i=1}^{2j}R_{i}}{g_{s}\left( \alpha
^{\prime }\right) ^{\frac{2j+1}{2}}}\left( \prod_{i=1}^{l}\left[ 1+\left(
2\pi \alpha ^{\prime }B_{i}\right) ^{2}\right] ^{\frac{1}{2}}\right) ;
\label{nbpsmass}
\end{equation}
while the mass of D$0$-branes is 
\begin{equation}
M_{\text{D}0}=\sqrt{2}\frac{1}{g_{s}\left( \alpha ^{\prime }\right) ^{\frac{1%
}{2}}}.
\end{equation}
In eq (\ref{nbpsmass}) $B_{i}$ denotes the $\left( 2i-1,2i\right) $
components of NS-NS $B_{\mu \nu }$ field which, in terms of $\theta _{i}$,
is given by 
\begin{equation}
B_{i}=\frac{1}{2\pi R_{2i+1}R_{2i}\theta _{i}}.
\end{equation}

Note that the relations (\ref{nbpsmass}) may be derived from the relation $%
M_{\text{D}2j}=G_{s}C_{\text{D}2j}$ and the identity \cite{i} 
\begin{equation}
G_{s}=g_{s}(\frac{\text{det}\left( g+2\pi \alpha ^{\prime }B\right) }{\text{%
det}B})^{\frac{1}{2}}
\end{equation}
where $g_{s}$\ is the closed string coupling constant.

Note also that in large non-commutativity, the $M_{\text{D}2j}$ masses can
be linked with the mass $M_{\text{D}2l}$ of the original brane. Indeed,
taking the large limits of all $B_{i}$'s in eq (\ref{nbpsmass}) one gets 
\begin{equation}
M_{\text{D}2l}=\sqrt{2}\frac{1}{\left( \prod_{i}^{l}\theta _{i}\right)
g_{s}\left( \alpha ^{\prime }\right) ^{\frac{1}{2}}}=\frac{1}{\left(
\prod_{i}^{l}\theta _{i}\right) }M_{\text{D}0}.
\end{equation}
If instead of taking all the $B_{i}$'s large, we keep one of them, say $%
B_{n} $ finite, one finds 
\begin{equation}
M_{\text{D}2l}=\frac{1}{\left( \prod_{i\neq n}^{l}\theta _{i}\right) }M_{%
\text{D}2}^{\left( n\right) }.
\end{equation}
with 
\begin{equation}
M_{\text{D}2}^{\left( n\right) }=\sqrt{2}\frac{R_{2n-1}R_{2n}}{g_{s}\left(
\alpha ^{\prime }\right) ^{\frac{3}{2}}}\left[ 1+\left( 2\pi \alpha ^{\prime
}B_{n}\right) ^{2}\right] ^{\frac{1}{2}}.
\end{equation}

More generally taking $B_{i_{1}},B_{i_{2,}},...,B_{i_{s}}$ large and $%
B_{i_{s+1}},B_{i_{_{s+2},}},...,B_{i_{l}}$ $\left( 1\leq s\leq l\right) $
finite\ and putting back in eq (\ref{nbpsmass}), one finds the following
D-branes mass relations 
\begin{equation}
M_{\text{D}2l}=\frac{1}{\left( \prod_{r=1}^{s}\theta _{i_{r}}\right) }M_{%
\text{D}2s}^{\left( n\right) }.
\end{equation}

Substituting these relations into the developement (\ref{energydevlop}), one
gets an energy formula $E_{\text{total}}$ expanded in terms of the masses $M_{%
\text{D}2j}$ of D$2j$-branes $\left( 1\leq j\leq l\right) $%
\begin{equation}
E_{\text{total}}=NM_{\text{D}2l}+\sum_{i=1}^{l}N_{i}M_{\text{D}2}^{\left(
i\right) }+\sum_{i,j=1}^{l}N_{ij}M_{\text{D}4}^{\left( ij\right) }+\ldots
+N_{i_{1}\ldots i_{l}}M_{\text{D}0}.  \label{totaldevlop}
\end{equation}

In this stage one may ask what does this formula mean? As the energy is\
upper bounded by $M_{\text{D}2l}$ $\left( E_{\text{total}}\leq M_{\text{D}%
2l}\right) $, it seems that the original unstable $M_{\text{D}2l}$
annihilates to different kinds of $M_{\text{D}2j}$ branes $\left( 0\leq
j\leq l\right) $ in the vacuum. To interpret this spectrum; let us first
consider the case of a non-BPS D2-brane on $\mathbb{T}_{\theta }^{2}$. In
this case, which corresponds to $l=1$ in eq (\ref{totalmass}), the energy eq(%
\ref{totaldevlop}) splits as follows 
\begin{equation}
E\left( \mathcal{P}_{n+m\theta }\right) =nM_{\text{D}2}+mM_{\text{D}0}.
\label{specialenergy}
\end{equation}
Following \cite{p}, it is natural to interpret this spectrum as a bound
state of $n$ D2-branes and $m$ D0-branes. This interpretation which is
dictated by the study of tachyon condensation is also supported by T-duality
and the analysis of the\ following exact mass spectrum of the \{$m$D0, $n$%
D2\} bound state 
\begin{equation}
M_{\left( n,m\right) }=\sqrt{2}\frac{R_{1}R_{2}}{g_{s}\left( \alpha ^{\prime
}\right) ^{\frac{3}{2}}}\left[ 1+\left( 2\pi \alpha ^{\prime }B_{\text{eff}%
}\right) ^{2}\right] ^{\frac{1}{2}},  \label{dmass}
\end{equation}
where $B_{\text{eff}}$ is an effective field, including the flux due to $m$
D0-branes, 
\begin{equation}
B_{\text{eff}}=B+\frac{1}{2\pi R_{1}R_{2}}\frac{m}{n}.
\end{equation}
Taking the\ large limit of $\left( 2\pi \alpha ^{\prime }B_{\text{eff}%
}\right) $, then the mass formula (\ref{specialenergy}) is reproduced.
Extending this study to the problem of non-BPS D$2l$-branes on the
non-commutative torus $\mathbb{T}_{\mathbf{\theta }}^{2l}$, our mass formula
(\ref{totaldevlop}) may be understood as describing bound states of $N$D$2l$%
-branes, $\left( \sum_{i=1}^{l}N_{i}\right) $D$\left( 2l-2\right) $-branes, $%
\left( \sum_{i,j=1}^{l}N_{ij}\right) $D$\left( 2l-4\right) $-branes,$\cdots $%
, $\left( \sum_{i_{1}\ldots i_{s}}^{l}N_{i_{1}\ldots i_{s}}\right) $D$(2l-$ $%
s)$-branes,$\cdots $, $\left( \sum_{i_{1}\ldots i_{s}}^{l}N_{i_{1}\ldots
i_{l-1}}\right) $D$2$-branes and finally $\left( \prod_{i=1}^{l}m_{i}\right) 
$ D$0$-branes. Note that if one is considering the general situation where
all the $2l$ radii $R_{2i-1}$ and $R_{2i}$ of the $2l$-dimensional torus\
are different, then one should distinguish various kinds of D$2j$-branes for
a given $j$ $\left( 1\leq j<l\right) $.

\bigskip

\begin{center}
{\small \bigskip } 
\begin{tabular}{|c|c|c|c|c|}
\hline
{\small D2j branes} & {\small mass} & {\small \# of } & {\small Kind of } & 
{\small Total \# of } \\ 
{\small \ in the bs} &  & {\small D2j branes} & {\small D2j branes} & 
{\small D2j branes} \\ \hline
{\small D0} & $\frac{\sqrt{2}}{g_{s}\left( \alpha ^{\prime }\right) ^{1/2}}$
& $m_{1}m_{2}$ & {\small 1} & $m_{1}m_{2}$ \\ \hline
{\small D}$^{\left( 1\right) }${\small 2} & $\frac{\sqrt{2}R_{1}R_{2}}{%
g_{s}\left( \alpha ^{\prime }\right) ^{3/2}}\Omega _{1}$ & $n_{1}m_{2}$ & 2
& $n_{1}m_{2}+n_{2}m_{1}$ \\ \cline{2-3}
{\small D}$^{\left( 2\right) }${\small 2} & $\frac{\sqrt{2}R_{3}R_{4}}{%
g_{s}\left( \alpha ^{\prime }\right) ^{3/2}}\Omega _{2}$ & $n_{2}m_{1}$ &  & 
\\ \hline
{\small D4} & $\frac{\sqrt{2}R_{1}R_{2}R_{3}R_{4}}{g_{s}\left( \alpha
^{\prime }\right) ^{5/2}}\Omega _{1}\Omega _{2}$ & $n_{1}n_{2}$ & {\small 1}
& $n_{1}n_{2}$ \\ \hline
\end{tabular}
\end{center}

\vspace{0,5cm}

\begin{center}
Table (b)
\end{center}

In table (b) we have given the complete spectrum of the non-BPS branes
involved in the bound state (bs) \{D$4$,D$2$,D$0$\}. The general spectrum of
\{D$2l$,D($2l-2$),$\ldots $,D$2$,D$0$\} bound state for all $\theta _{i}$'s
irrational is given in table (c).

\bigskip

\begin{center}
\begin{tabular}{|c|c|c|c|c|}
\hline
{\small D2j branes} & {\small mass} & {\small \# of } & {\small Kind of } & 
{\small Total \# of } \\ 
{\small \ in the bs} &  & {\small D2j branes} & {\small D2j branes} & 
{\small D2j branes} \\ \hline
{\small D0} & $\frac{\sqrt{2}}{g_{s}\left( \alpha ^{\prime }\right) ^{1/2}}$
& {\small M=}$\Pi _{i=1}^{l}m_{i}$ & {\small 1} & $\Pi _{i=1}^{l}m_{i}$ \\ 
\hline
{\small D}$^{\left( i\right) }${\small 2} & $\frac{\sqrt{2}R_{2i-1}R_{2i}}{%
g_{s}\left( \alpha ^{\prime }\right) ^{3/2}}\Omega _{i}$ & $\frac{n_{i}}{%
m_{i}}M$ & $l$ & $n_{1}m_{2}+n_{2}m_{1}$ \\ \hline
{\small D}$^{\left( ij\right) }${\small 4} & $\frac{\sqrt{2}%
(R_{2i-1}R_{2i}\Omega _{i})(R_{2j-1}R_{2j}\Omega _{j})}{g_{s}\left( \alpha
^{\prime }\right) ^{5/2}}$ & $\frac{n_{i}n_{j}}{m_{i}m_{j}}M$ & $\frac{%
l\left( l-1\right) }{2}$ & $M\Sigma _{ij}\frac{n_{i}n_{j}}{m_{i}m_{j}}$ \\ 
\hline
$\cdots $ & $\cdots $ & $\cdots $ & $\cdots $ & $\cdots $ \\ \hline
{\small D}$^{\left\{ i_{1},..,i_{s}\right\} }${\small 2}$s$ & $\frac{\sqrt{2}%
\Pi _{i=1}^{s}(R_{2i_{s}-1}R_{2i_{s}}\Omega _{i_{s}})}{g_{s}\left( \alpha
^{\prime }\right) ^{\left( 2s+1\right) /2}}$ & $M\Pi _{r=1}^{s}(\frac{n_{ir}%
}{m_{ir}})$ & $\frac{l!}{\left( l-s\right) !s!}$ & $M\Sigma \{\Pi _{r=1}^{s}(%
\frac{n_{ir}}{m_{ir}})\}$ \\ \hline
$\cdots $ & $\cdots $ & $\cdots $ & $\cdots $ & $\cdots $ \\ \hline
{\small D2}$l$ & $\frac{\sqrt{2}\Pi _{i=1}^{l}(R_{2i-1}R_{2i}\Omega _{i})}{%
g_{s}\left( \alpha ^{\prime }\right) ^{\left( 2l+1\right) /2}}$ & $\Pi
_{i=1}^{l}n_{i}$ & {\small 1} & $\Pi _{i=1}^{l}n_{i}$ \\ \hline
\end{tabular}
\end{center}

\vspace{0,5cm}

\begin{center}
Table (c)
\end{center}

Let us make two comments concerning this spectrum.

\begin{description}
\item 
\begin{itemize}
\item  Inspired from the D$0$-D$2$ bound state analysis, we can derive the
exact mass formula $M_{\left\{ n_{1},\ldots ,n_{l},m_{1},\ldots
,m_{l}\right\} }$ of the \{D$2l$,D$(2l-2)$,$\ldots $,D$2$,D$0$\} bound
states. We show that 
\begin{equation}
M_{\left\{ \mathbf{n},\mathbf{m}\right\} }=\sqrt{2}\frac{\prod_{i=1}^{l}%
\left( R_{2i-1}R_{2i}n_{i}\right) }{g_{s}\left( \alpha ^{\prime }\right) ^{%
\frac{2l+1}{2}}}\left( \prod_{j=1}^{l}\left[ 1+\left( 2\pi \alpha ^{\prime
}B_{i\text{ eff}}\right) ^{2}\right] ^{\frac{1}{2}}\right) ,
\end{equation}
where$\ \mathbf{n}=(n_{1},\ldots ,n_{l})$, $\mathbf{m=(}m_{1},\ldots ,m_{l})$
and$\ B_{\text{eff}}$ is an effective field given by 
\begin{equation}
B_{i\text{,eff}}=B_{i}+\frac{1}{2\pi R_{2i-1}R_{2i}}\frac{m_{i}}{n_{i}}%
;\quad i=1,\cdots ,l.
\end{equation}
\end{itemize}
\end{description}

\begin{itemize}
\item  In the case where some of the $\mathbb{T}_{\theta _{i}}^{2}$\ factors
of the non-commutative $2l$ dimensional torus are fuzzy torii, that is some
of the $\theta _{i}$'s, $1\leq i\leq s$, are irrational and the remaining
ones are rational; $\theta _{i}=q_{i}/p_{i}$ for $s+1\leq i\leq l$, one
still has bound states of type \{D$2s$,D$2s-2$,$\ldots $,D$2$,D$0$\} whose
spectrum mass may be read from table (c) in addition to extra D$0$-branes.
In this case the tachyon soliton $T$ corresponding to such representation is 
\begin{equation}
T=t_{\text{max}}.\left( \overset{s}{\underset{i=1}{\otimes }}\mathcal{P}%
_{\left\{ n_{i}+m_{i}\theta _{i}\right\} }\right) \otimes \left( \overset{l}{%
\underset{i=s+1}{\otimes }}\Pi _{k_{i}}\right) .  \label{mixedproj}
\end{equation}
For level $\mathbf{k}_{s}\mathbf{=(}k_{i_{s+1}},...,k_{l})$ vacuum
configurations on the fuzzy torus, the energy of\ the solitons is given by 
\begin{equation}
E\left( \mathcal{P}_{\left\{ n_{i}+m_{i}\theta _{i}\right\} },\Pi
_{k_{i}}\right) =\left( \prod_{i=1}^{s}\left( n_{i}+m_{i}\theta _{i}\right)
\right) \left( \prod_{i=s+1}^{l}\frac{k_{i}}{p_{i}}\right) .
\label{mixedenergy}
\end{equation}
\end{itemize}

In what follows we\ want to show that the above configurations (\ref
{mixedproj},\ref{mixedenergy}) do not have the same bound states\ and possed
the following goup $G$ as a full gauge symmetry 
\begin{equation}
G=U\left( \prod_{i=1}^{s}k_{i}^{(ir)}\right) \times U\left(
\prod_{i=s+1}^{l}k_{i}^{(r)}\right) .
\end{equation}
For bound states, this is clearly seen on the eq(\ref{mixedenergy}) since
they appear from the Powers-Rieffel projectors in one to one correspondence
with irrational $\theta _{i}$'s. For a fixed value of $s$, $\left( 1\leq
s\leq l\right) $, one has bound states of type \{D$2s$,D$(2s-2)$,$\ldots $,D$%
2$,D$0$\} whose spectrum varies with $s$. Concerning the gauge symmetry, it
is interesting to note first that for $s=0$, $(h=0)$, that is all $\theta
_{i}$'s are rational. one is in presence of only D$0$-branes and so the
gauge symmetry is $U\left( \prod_{i=1}^{l}k_{i}^{(r)}\right) $. For non-zero 
$s$, the situation is a little bit subtle since there exists bound states
involving other kinds of D$2j$-branes for $0\leq j\leq s$ in addition to the
D$0$-branes of the rational factors having $U\left(
\prod_{i=s+1}^{l}k_{i}^{(r)}\right) $ as a gauge group. The $U\left(
\prod_{i=1}^{s}k_{i}^{(ir)}\right) \times \prod_{h=1}^{l}U\left(
N_{i_{1}i_{2}}^{(ir)}.._{.i_{h}}\right) $ gauge group of the\ D$2j$-branes
for $0\leq j\leq s$ system may be obtained from symmetry of the
non-commutative $2l$-dimensional torus. Using the T-duality transformations 
\begin{eqnarray}
\theta _{i}^{\prime } &=&\frac{\alpha _{i}-\beta _{i}\theta _{i}}{\gamma
_{i}-\delta _{i}\theta _{i}};i=1,\ldots ,l \\
\mathfrak{A}_{\theta _{i}^{\prime }} &\sim &\mathfrak{A}_{\theta _{i}},\quad \mathfrak{A}%
_{\mathcal{\theta }^{\prime }}\sim \mathfrak{A}_{\mathbf{\theta }},
\end{eqnarray}
with $\alpha _{i}\delta _{i}-\beta _{i}\gamma _{i}=1$ and $\alpha _{i},$ $%
\beta _{i},$ $\gamma _{i},$ $\delta _{i}\in \mathbb{Z}$; leaving the total
energy (\ref{totaldevlop}) invariant.\ Under the above $\left[ SL(2,\mathbb{Z%
})\right] ^{l}$ dualities, it is usually possible to rewrite eqs (\ref
{mixedenergy}) as 
\begin{equation}
E(\mathcal{P}_{\left\{ n_{i}+m_{i}\theta _{i}\right\} },\Pi _{k_{i}})=\left[
\prod_{i=1}^{s}k_{i}^{(ir)}\left( \beta _{i}+\alpha _{i}\theta _{i}\right) %
\right] \left( \prod_{i=s+1}^{l}\frac{k_{i}^{(r)}}{p_{i}}\right) ;
\label{tmixedenergy}
\end{equation}
where the $k_{i}$'s $(1\leq i\leq s)$ are the greatest commun divisor of $%
\left( n_{i},m_{i}\right) $ pairs. In the case of the D$2$-D$0$ bound state; 
$\left( n+m\theta \right) M_{\text{D2}}$ was interpreted in \cite{p} as the
energy of $k$ bound states of D$0$-branes with a $U(k)$ gauge symmetry.
Extending this reasoning to the non-commutative torus $\mathbb{T}_{\mathbf{%
\theta }}^{2l}$ and using eq (\ref{tmixedenergy}), the total number of
D0-branes is $\prod_{i=1}^{s}\left( k_{i}^{(ir)}\right) $ coming from the
irrational case and $\prod_{i=s+1}^{l}\left( k_{i}^{(r)}\right) $ from the
rational one. Seen as two different sets of D0-branes, one concludes that
the gauge symmetry is $\prod_{i=1}^{s}\left( k_{i}^{(ir)}\right) \times
\prod_{i=s+1}^{l}\left( k_{i}^{(r)}\right) $. If we assume that all
D0-branes of the soliton are indistinguishable, then we end with a larger
gauge group $\prod_{i=1}^{l}\left( k_{i}\right) $ containing the previous
symmetries as subgroups.

\section{Discussion and Conclusion}

In this paper we have studied the solitonic solutions of a non-BPS D$2l$%
-branes; $l\geq 1$; on both Moyal spaces $R_{\mathbf{\theta }}^{2l}$ and
non-commutative torii $T_{\mathbf{\theta }}^{2l}$. Actually this study may\
also be viewed as an extension of the analysis of the tachyon condensation
made in \cite{p} for non-BPS D$2$-branes in presence of a constant NS-NS $B$%
-field. Our results may be summarised into the following two:

(1) We have derived soliton solutions for non-BPS D$2l$-branes; $l\geq 1;$
on $R_{\mathbf{\theta }}^{2l}$. In particular, we have shown that besides
the usual stable vacuum field configuration, there exists also quasi-stable
solutions minimising the total energy $E_{\text{total}}=E(\theta _{1},\theta
_{2},...,\theta _{l}).$ These solutions are associated with local minima of $%
E_{\text{total}}$.\ and are interpreted as non-BPS D$2s_{1}$-branes, $0\leq
s_{1}\leq l-1$, representing the $l$ decay levels of the original non-BPS $%
D2l$-brane into lower dimensional world volume branes. Besides the original
non-BPS D$2l$-brane and the non-BPS D$0$-branes one ends up with $2^{l}-2$
brane configurations partionned as 
\begin{eqnarray}
\lbrack D2l] &\backsim &\sum_{s_{1}=1}^{l-1}\frac{l!}{(l-s_{1})!s_{1}!}[%
D2s_{1}]  \notag \\
&=&l[D2]+\frac{l(l-1)}{2}[D4]+\frac{l(l-1)(l-2)}{6}[D6]+...  \notag \\
&&+\frac{l(l-1)(l-2)}{6}[D(2l-6)]+\frac{l(l-1)}{2}[D(2l-4)]+l[D(2l-2)].
\end{eqnarray}
This expansion means that at a given stage of the condensation, say at a
step $j$, $\left( 0<j<l\right) $, one has  $\frac{l!}{(l-j)!j!}$ kinds of D$%
2j$-branes. For $j=1$ and $j=l-1$ for instance, we have respectively $l$
world volumes D$2$-branes and D$(2l-2)$-branes and for $j=2$\ and $j=l-2$ we
have $\frac{l(l-1)}{2}$ world volumes $D4$-branes and a similar number of $%
D(2l-4)$-branes. In figure \ref{fig:cur} we have proposed a shape of the\ $%
V(\phi )$ potential with $l$ waves to describe these quasi-stable solutions.

Moreover, since for a generic extremum $s_{1}$, the corresponding D$2s_{1}$%
-brane, $1\leq s_{1}\leq l-1$ is an unstable configuration, one can imagine
that D$2s_{1}$\ itself condensates into lower dimensional world volumes D$%
2s_{2}$-branes, $0\leq s_{2}\leq s_{1}-1$. Thus, given an integer $s_{2}$,
we have here also \ 
\begin{equation*}
\lbrack D2s_{1}]-\backsim \sum_{s_{2}=1}^{s_{1}-1}\frac{s_{1}!}{%
(s_{1}-s_{2})!s_{2}!}[D2s_{2}].
\end{equation*}
More generally we have the following result: Starting from the original
non-BPS D$2l$-brane on $R_{\mathbf{\theta }}^{2l}$; with $l\geq 1$; and
taking the weaker definition of large non commutativity which ammounts to
put one of the $\theta _{i}$' s large and all remaining others finite, one
gets $l$ kinds of $(2l-2)$-dimensional solitons identified with D$(2l-2)$%
-branes. Repeating the same mechanism to each one of the $l$ D$(2l-2)$%
-branes, one obtains, after integrating out equivalent configurations, $%
\frac{l(l-1)}{2}$ D$(2l-4)$-branes on the $R_{\mathbf{\theta }}^{2(l-2)}$
Moyal space. Successive iterations lead at the end to $\Pi _{s=0}^{l}(\frac{%
l!}{(l-s)!s!})$ D$0$-branes. Furthermore if we denote by $k_{s}$, the level
of the D$2s$ solitons with $k_{s}\geq 1$ and\ $k_{l}=1$\ one gets $U(\Pi
_{s=0}^{l}[\frac{k_{s}l!}{(l-s)!s!}])$ as a\ full gauge symmetry.

(2)\ We have studied the tachyon solitons on $2l$-dimensional
non-commutative torii using both rational and irrational representations. We
have determined the solitons mass spectrum and shown for $s$ irrational $%
\theta _{i}$'s, $1\leq s\leq l$,\ the existence of general bound states \{D$%
2s$, D$(2s-2)$,$...$,D$2i$,$...$, D$2$, D$0\},$ extending the D$2$-D$0$
bound state obtained in \cite{p} for the case of a non-BPS D$2$-brane on an
irrational two torus. From the mass formula of our bound state, namely 
\begin{equation}
M_{\left\{ \mathbf{n}_{s},\mathbf{m}_{s}\right\} }=\sqrt{2}\frac{%
\prod_{i=1}^{s}\left( R_{2i-1}R_{2i}n_{i}\right) }{g_{s}\left( \alpha
^{\prime }\right) ^{\frac{2l+1}{2}}}\left( \prod_{j=1}^{s}\left[ 1+\left(
2\pi \alpha ^{\prime }B_{i\text{ eff}}\right) ^{2}\right] ^{\frac{1}{2}%
}\right) ,
\end{equation}
with $\mathbf{n}_{s}=(n_{1},n_{2}...,n_{s})$ and $\mathbf{m}%
_{s}=(m_{1},m_{2}...,m_{s})$, or equivalently by using eq(\ref{dmass}) 
\begin{equation}
M_{\left\{ \mathbf{n},\mathbf{m}\right\} }=\prod_{i=1}^{s}\{M_{\left(
n_{i},m_{i}\right) }\},
\end{equation}
one learns that\ \{$D2s,D(2s-2),...,D2i,...,D2,D0\}$\ system is unstable and
decays into $s$ \{$D2-D0\}^{(i)}$ bound states of masses $M_{\left(
n_{i},m_{i}\right) }.$\ Note that for the case $s=0$; i.e. no\ $\theta _{i}$
is irrational, we have $D0$-branes but no bound state as expected. For $s=l$%
; i.e. all\ $\theta _{i}^{,}$ s are irrational, we have bound states type \{$%
D2l,D(2l-2),...,D2s,...,D2,D0\}$ which as suggested desintegrate into $l$ \{$%
D2-D0\}$ bound states.\ Applying the Bars\textit{\ et al} analysis made for
the $D2-D0$ \ bound state to the \{$D2s,D(2s-2),...,D2i,...,D2,D0\}$\
system, it is no difficult to check that the full gauge group of the vacuum
configuration is $U(\Pi _{i=1}^{s}k_{i}^{(ir)}\Pi _{i=s+1}^{l}k_{i}^{(r)})$.
It contains as a subgroup $U(\Pi _{i=1}^{s}k_{i}^{(ir)})\times U(\Pi
_{i=s+1}^{l}k_{i}^{(r)}),$\ the gauge symmetry coming from the irrational
and rational sectors.\ The potential describing solitons on \ $T_{\mathbf{%
\theta }}^{2l}$ has a similar shape as in figure \ref{fig:cur} except that one has now
to specify:

\begin{itemize}
\item  The representation considered for $T_{\mathbf{\theta }}^{2l}$:
rational or irrational. 

\item  The interpretation of the extrema in terms of bound states. Extension
of these results to other cases such as\ non-commutative orbifods will
considered in a future occasion.\bigskip 
\end{itemize}

\begin{center}
{\huge Acknowlegement}
\end{center}

This research work has been supported by SARS, ``programme de soutien \`{a} la
recherche scientifique de l'universit\'{e} Mohammed V-Agdal, Rabat.'' 


\newpage

\begin{thebibliography}{10}

\bibitem[1]{a}  A. Sen, ``Tachyon condensation on the brane anti-brane system
,'' \textbf{JHEP 9808:012}(1998), hep-th/9805170.

\bibitem[2]{b}  A. Sen, ``SO(32) spinors of type I and other solitons on
brane anti-brane pair,'' \textbf{JHEP 9809:023}(1998), hep-th/9808141.

\bibitem[3]{c}  A. Sen, ``Stable non-BPS bound states of BPS D-branes,'' 
\textbf{JHEP 9808:010}(1998), hep-th/9912249.

\bibitem[4]{d}  N. Berkovits, A. Sen and B. Zwiebach; ``Tachyon condensation
in superstring field theory,'' hep-th/0002211.

\bibitem[5]{e}  A. Sen and B. Zwiebach; ``Tachyon condensation in string
field theory,'' hep-th/9912249.

\bibitem[6]{f}  E. Witten, ``Noncommutative geometry and string field
theory,'' Nucl: Phys. \textbf{B268}, 253 (1986)

\bibitem[7]{g}  A. Connes, M.R. Douglas and A. Schwarz,\textbf{JHEP 9802:003 
}(1998), hep-th/9711162

\bibitem[8]{h}  M.R. Douglas and C. Hull, \textbf{JHEP 9802:008 }(1998),
hep-th/9711165

\bibitem[9]{i}  N. Seiberg and E. Witten, ``String theory and
non-commutative geometry,'' \textbf{JHEP 9909:032} (1999), hep-th/9908142.

\bibitem[10]{j}  R. Gopakumar, S. Minwalla and A.
Strominger,``Noncommutative solitons,'' hep-th/0003160.

\bibitem[11]{k}  J. A. Harvey, P. Kraus, F. Larsen and E. J. Martinec,
``D-branes and strings as noncommutative solitons,'' hep-th/0005031.

\bibitem[12]{l}  J. A. Harvey, P. Kraus and F. Larsen, ``Exact
noncommutative solitons,'' hep-th/0010060.

\bibitem[13]{m}  D. V. Gross and N. A. Nekrasov, ``Monopoles and strings in
noncommutative gauge theory,'' hep-th/0005204.

\bibitem[14]{n}  D. V. Gross and N. A. Nekrasov, ``Dynamics of strings in
noncommutative gauge theory,'' hep-th/0007204.

\bibitem[15]{o}  D. V. Gross and N. A. Nekrasov, ``Solitons in
noncommutative gauge theory,'' hep-th/0010090.

\bibitem[16]{p}  I. Bars, H. Kajiura, Y. Matsuo and T. Takayanagi, ``Tachyon
condensation on non-commutative torus,'' hep-th/0010101.

\bibitem[17]{q}  E. Witten, ``Noncommutative tachyons and string field
theory,'' hep-th/0006071.

\bibitem[18]{r}  W. Nahm, Phys. Lett. \textbf{90B} (1980) 413

\bibitem[19]{s}  W. Nahm,"The Constraction of all Self-dual Multimonopoles by the ADHM Method'', in
"Monopoles in quantum field theory,'' Craigie \textit{et al.} Eds., Wrold
Scientific, Singapore (1982);

\bibitem[20]{t}  N. J. Hitchin, Comm. Math. Phys. \textbf{89} (1983) 145.
\end{thebibliography}
\end{document}